\documentclass[review]{elsarticle}
\makeatletter
\def\ps@pprintTitle{%
	\let\@oddhead\@empty
	\let\@evenhead\@empty
	\def\@oddfoot{\centerline{\thepage}}%
	\let\@evenfoot\@oddfoot}
\makeatother

\usepackage{lineno,hyperref}
\modulolinenumbers[5]
\usepackage{booktabs}
\usepackage{graphicx}
\usepackage{color}
\usepackage{bm}

\usepackage{amsmath,amssymb,array}
\usepackage{amsthm}
\usepackage[utf8]{inputenc}
\usepackage[T1]{fontenc}
 \usepackage{relsize}
\usepackage{caption}
\usepackage{subcaption}

\begin{document}

\begin{frontmatter}

\title{qGaussian: Tools to Explore Applications of Tsallis Statistics\\}

\author[mymainaddress]{W. S.de Lima}
\author[mymainaddress]{E.L de Santa Helena \corref{mycorrespondingauthor}}
\ead{elsh@ufs.br}

\address[mymainaddress]{Departamento de F\'{\i}sica, Universidade Federal do Sergipe, Aracaju, Brazil.}


\begin{abstract}

$q$-Gaussian distribution appear in many science areas where we can find systems that could be described within a nonextensive framework.
	Usually, a way to assert that these systems belongs to nonextensive framework is by means of numerical data analysis.
	To this end, we implement random number generator for $q$-Gaussian distribution, while we present how to computing its probability density function, cumulative density function and quantile function besides a tail weight measurement using robust statistics.
\end{abstract}

\end{frontmatter}


\section{Introduction}
Entropy is a fundamental concept in physics since it is in essence of second law of thermodynamics. In the article 'Possible generalization of Boltzmann-Gibbs statistics' \cite{T88}, was 
postulated the Tsallis entropy $S_q[p(x)]=(q-1)^{-1}*(1-\int_{-\infty}^{\infty}p^q(x)dx)$.
Within the nonextensive approach, the sum of the entropy of two independent subsystems is given by: $S_q(a+b)=S_q(a)+S_q(b)+(1-q)S_q(a)S_q(b)$, where $q$ is an entropic index.
The $q$-Gaussian
density function $p(x)$, presented in the next section, arises from maximizing the $q$-entropy functional $S_q[p(x)]$ under constraints \cite{T09} and \cite{H14} and it is important at the framework of the nonextensive statistical mechanics.

There is a broad literature where the nonextensive approach is used to model systems and/or explain many-body problems and 
issues related to chaos. In some cases, there exist strong evidences that theory works, certified by a broad numerical decade from experimental measurements or numericals simulation that let us obtain a $q$ value by an almost flawless curve fitting. In other cases, a lot of observational data are only suggestive of a nonextensive approach. A myriad of examples can be found in \cite{GT04} and \cite{T09}. This work aims to spread among users of R, a statistical package that deals with the distribution of $q$-Gaussian, allowing researchers to infer and evaluate, from empirical data, a nonextensive behaviour.

At section theoretical background, we first see a way to represent the probability density function and how to write the cumulative density function and quantile function using the Beta function. Then, we present the random number generator and a way to identify the $q$-Gaussian at empirical data. Next, we will see the implementation in R of all subjects described previously. Lastly, illustrative examples section present the density function shape, after the  comparison among $q$-Gaussian with its special cases and an estimate of $q$ value is made up from a data set. 

\section{Theoretical background}

The $q$-Gaussian probability density function, named here $q$PDF \cite{SH15}, with $q$-mean $\mu_q$ and $q$-variance $\sigma_{q}$ can be written as:
\begin{equation}
p(x;\mu_q,\sigma_q)=\frac{1}{\sigma_q {\text B}\left(\frac{\alpha}{2},\frac{1}{2}\right)}
\sqrt{
\frac{|Z|}{u^{(1+1/Z)}}
}
\label{pdf}
\end{equation}
where  $Z=(q-1)/(3-q)$,
\begin{equation*}
\alpha=\left\{
\begin{array}{ll}
1-1/Z& \textrm{if  $q<1$}, \\
1/Z  & \textrm{if $1<q<3$}, \\
\end{array}\right.
\end{equation*}
 
$u(x)= 1+Z (x-\mu_q)^2 /  \sigma^{2}_{q}  $, and $\text B(a,b)$ is the  Beta function \footnote[1]{
	Beta function:
	$\text B(a,b) = \int_{0}^{1} t^{a-1}(1-t)^{b-1}dt.$
	}.
In the limit of $q\rightarrow 1$ a $q$PDF tends to a standard Gaussian distribution. 
For $q < 1$, it is a compact support distribution, with $x \in [ \pm\sigma_q / \sqrt{-Z}]$. When $1 < q < 3$,  it is a heavy tail. In the last case, a power law asymptotic behaviour describes well this class of distribution.

Given $X=(x_1,x_2,...x_n)$ a random variable, the cumulative distribution function ($q$CDF), $F(x)=P[X < x]$
\begin{equation*}\label{csum}
F(x) = \int_{-\infty}^{x}p(v)dv,
\end{equation*}
 could be represented through $\text I_{w}(a,b)$, the regularized incomplete Beta function
  \footnote[2]{
 	Incomplete Beta function:
 	$\text B_{w}(a,b) = \int_{0}^{w} t^{a-1}(1-t)^{b-1}dt$
 	
 	and the regularized incomplete Beta function: $\text I_{w}(a,b) = \text B_{w}(a,b)/\text B(a,b)$
 	}
  \cite{AS83}:

\begin{equation}
\label{cdf}
F(x;\mu_q;\sigma_q) = \left\{
\begin{array}{ll}
\frac{1}{2} \text I_{\beta}\left(\frac{\alpha}{2},\frac{1}{2}\right) & \textrm{if  $x<\mu_q$}, \\
1-\frac{1}{2} \text I_{\beta}\left(\frac{\alpha}{2},\frac{1}{2}\right) & \textrm{if  $x> \mu_q$}, \\
\end{array}\right.
\end{equation}
where
\begin{equation*}
\beta=\left\{
\begin{array}{ll}
u(x) & \textrm{if  $q<1$}, \\
1/u(x) & \textrm{if $1<q<3$}. \\
\end{array}\right.
\end{equation*}

In third, the quantile function ($q$QF) is obtained as an inverse function of $y=F(x)$, 
$w=\text I^{-1}_{2y}(\frac{\alpha}{2},\frac{1}{2})$: 

\begin{equation}\label{qdf}
C(y;\mu_q,\sigma_q) = \left\{
\begin{array}{ll}
\mu_q-\sigma_q^2 \sqrt{\left( \gamma -1\right )  /Z} & \textrm{if  $y<1/2$}, \\
\mu_q + \sigma_q^2 \sqrt{\left( \gamma -1\right )  /Z} & \textrm{if $1-y<1/2$}, \\
\end{array}\right.
\end{equation}
where
\begin{equation*}
\gamma=\left\{
\begin{array}{ll}
\text I^{-1}_{2y}\left(\frac{\alpha}{2},\frac{1}{2}\right) & \textrm{if  $q<1$}, \\
1/\text I^{-1}_{2y}\left(\frac{\alpha}{2},\frac{1}{2}\right) & \textrm{if $1<q<3$}. \\
\end{array}\right.
\end{equation*}

It is worth calling attention
despite the fact that $q$CDF and $q$QF obtained from a compact support distribution do not appear explicitly shown in  \cite{H14}, these could be deduced by the same straightforward method presented in it.

The random numbers generator from a $q$PDF 
can be implemented in different ways.
The straightforward method use the quantile function to
creates random sample elements $x_i=C(y_i;\mu_q,\sigma_q)$, where $y_i \in [0,1]$ are obtained from a uniform random number. In the second way, we use the Box-Muller algorithm as presented in \cite{TNT07} with the Mersenne-Twister algorithm as a uniform random number generator to create a $q$-Gaussian random variable $X\equiv N_q(\mu_q,\sigma_q) \equiv \mu_q+\sigma_q N_q(0,1)$ 
where $N_q(0,1)$ is called standard $q$-Gaussian.

The classical kurtosis methods, when applied to a data set, are very sensitive to outlying values, however, it is possible to diminish this problem at a
measurement of the tail heaviness by using robust statistic concepts. To doing that, given a sorted sample $\{x_1< \dots < \tilde{x}< \dots < x_n\}$ from a univariate distribution with median $\tilde x$,
 \cite{BHS06} established the medcouple   
to evaluates the tail weight  when applied to $\{x_{1} < \dots < \tilde{x}\}$ and 
$\{\tilde{x}< \dots < x_n\}$. This procedure can be applied to characterize the $q$-Gaussian with heavy tail and compact support. To this end, \cite{SH15} 
aiming to identify a $q$-Gaussian distribution at empirical data, proposed a relationship between medcouple and $q$ value obtained by curve fitting. 
  
\section{R implementation}

The main goal of the package \textbf{qGaussian} it is lets us to compute $q$PDF (\ref{pdf}), $q$CDF (\ref{cdf}) and $q$QF (\ref{qdf}) as same time generates random numbers from a $q$-Gaussian distribution parametrised by $q$ value. 
To compute the Beta function and its inverse it is necessary the 
\textbf{zipfR} package, while the \textbf{robustbase} is the package of robust statistic to implement a tail weight measurement.

\begin{table}[!h]
\begin{center}
\begin{tabular}{cc}
		\toprule
 Quantity &  R's commands   \\	
\midrule
$ p(x) $ & \fontfamily{cmss}\selectfont{ dqgauss(x,q,mu,sig) }  \\
$F(x) $  & \fontfamily{cmss}\selectfont{pqgauss(x,q,mu,sig,lower.tail=T)}   \\
$C(y)$ & \fontfamily{cmss}\selectfont{cqgauss(y,q,mu,sig,lower.tail=T) }    \\
$x_{i}$ & \fontfamily{cmss}\selectfont{rqgauss(n,q,mu,sig,meth="Box-Muller")} \\
$q $ & \fontfamily{cmss}\selectfont{qbymc(X)} \\
\bottomrule
\end{tabular}
\end{center}
\caption{Sintaxe of R's commands for each output quantities}
\label{tab}
\end{table}

In table \ref{tab}, the input argument {\fontfamily{cmss}\selectfont x} represent a vector of quantiles for instructions
{\fontfamily{cmss}\selectfont dqgauss(x,..)} 
and {\fontfamily{cmss}\selectfont pqgauss(x,..) } while {\fontfamily{cmss}\selectfont{y}} represent a vector of probabilities and {\fontfamily{cmss}\selectfont{n}} the length sample, for the random number generator. The parameters {\fontfamily{cmss}\selectfont{q}}, {\fontfamily{cmss}\selectfont{mu}} and {\fontfamily{cmss}\selectfont{sig}} are the entropic index, $q$-mean and $q$-variance, respectively,
assuming the default values $(0,0,1)$. The medcouple is used into the {\fontfamily{cmss}\selectfont{qbymc(X)}} code to estimate $q$ value and standard error, receiving a random variable {\fontfamily{cmss}\selectfont{X}}, from the class "vector",  as input.
We will see below, all the R's commands described above.

\section{Illustrative and demonstrative examples}

First of all, two packages should be loaded. \\
{\fontfamily{cmss}\selectfont
\noindent
library(robustbase) \\
\noindent
library(zipfR)
}

After, we start examples section presenting the shape of the $q$PDF plotted for typical $q$ values over a quantile range covering more that $99.9\%$ of area of the standard Gaussian. Besides that, we create a random sample with $q=0$ then we choose the  appropriate class intervals to create the histogram that is plotted against the $q$PDF. \\
 
{\fontfamily{cmss}\selectfont
\noindent
$\#\#\#$ Plot six qPDFs  \\
qv <- c(2.8, 2.5, 2, 1.01, 0, -5); nn <- 700  \\
xrg <- sqrt((3-qv[6])/(1-qv[6]))  \\
xr <- seq(-xrg, xrg, by = 2*xrg/nn)  \\
y0 <- dqgauss(xr, qv[6]) \\ 
plot(xr, y0, ty = 'l', xlim = range(-4.5, 4.5), ylab = 'p(x)', xlab = 'x') \\
for (i in 1:5){ \\
	if (qv[i] < 1) xrg <- sqrt((3-qv[i])/(1-qv[i]))  \\
	else xrg <- 4.5  \\
	vby <- 2*xrg/nn  \\
	xr <- seq(-xrg, xrg, by = 2*xrg/nn) \\
	y0 <- dqgauss(xr, qv[i]) \\
	points (xr, y0, ty = 'l', col = (i+1)) \\
}  \\
legend(2, 0.4, legend = c(expression(paste(q == -5)), expression(paste(q == 0)), \\
	expression(paste(q == 1.01)), expression(paste(q == 2)),  \\
	expression(paste(q == 2.5)), expression(paste(q == 2.8))),  \\
	col = c(1, 6, 5, 4, 3, 2), lty = c(1,1,1,1,1,1)  \\
	)  \\
	\\
$\#\#\#$ qPDF Histogram for q = 0  \\
qv <- 0  \\
rr <- rqgauss(2$\wedge$16, qv)  \\
nn <- 70  \\
xrg <- sqrt((3-qv)/(1-qv))  \\
vby <- 2*xrg/(nn)  \\
xr <- seq(-xrg, xrg, by = vby) \\
hist (rr, breaks = xr, freq = FALSE, xlab = "x", main = '') \\
y <- dqgauss(xr)  \\
lines(xr, y/sum(y*vby), cex = .5, col = 2, lty = 4)  \\
}

\begin{figure}[htbp]
	\centering
	\begin{subfigure}{.5\textwidth}
		\centering
		\includegraphics[width=1.\linewidth]{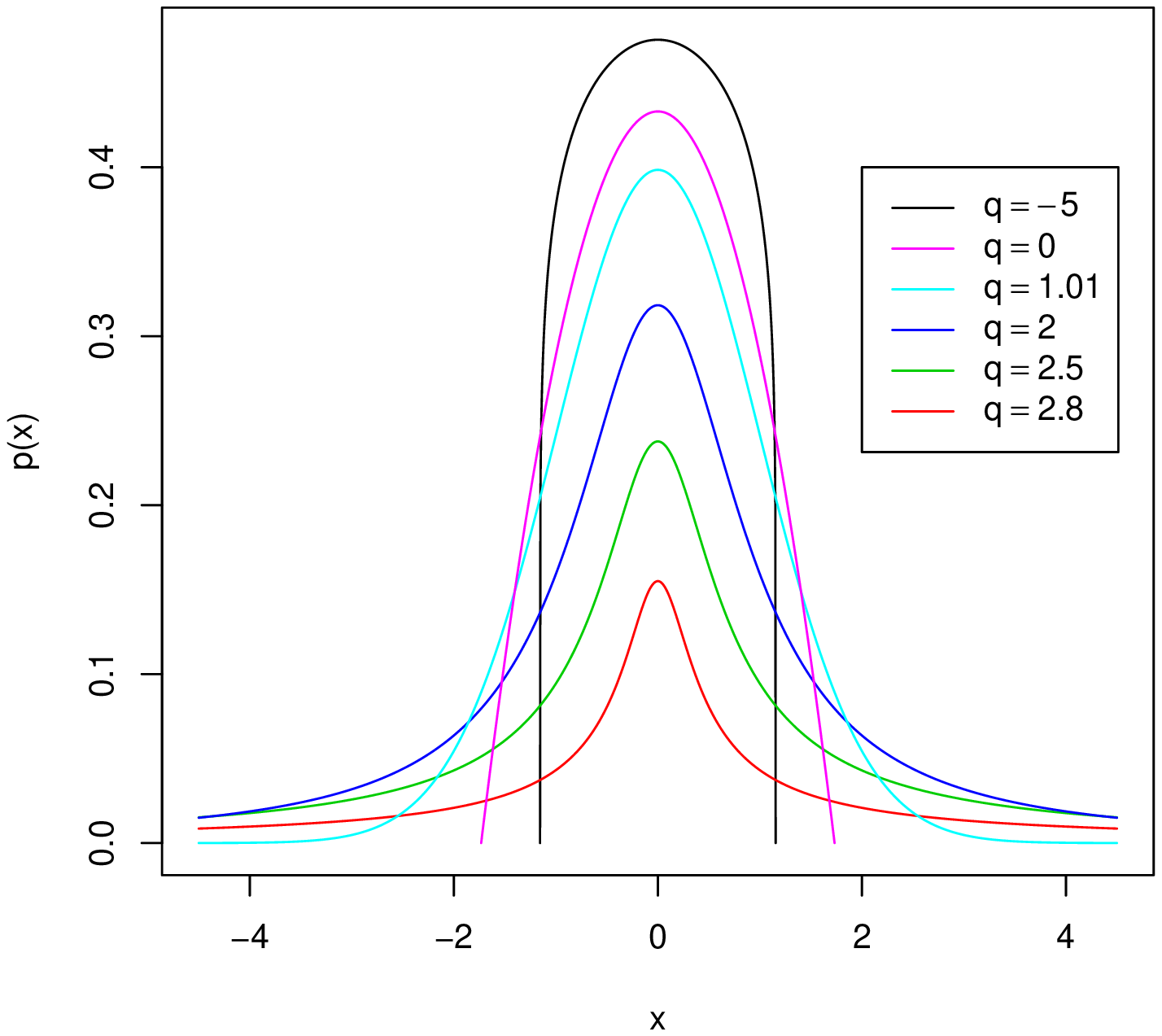}
		\caption{p(x) for typical $q$ values}
		\label{fig:sub1}
	\end{subfigure}%
	\begin{subfigure}{.5\textwidth}
		\centering
		\includegraphics[width=1.\linewidth]{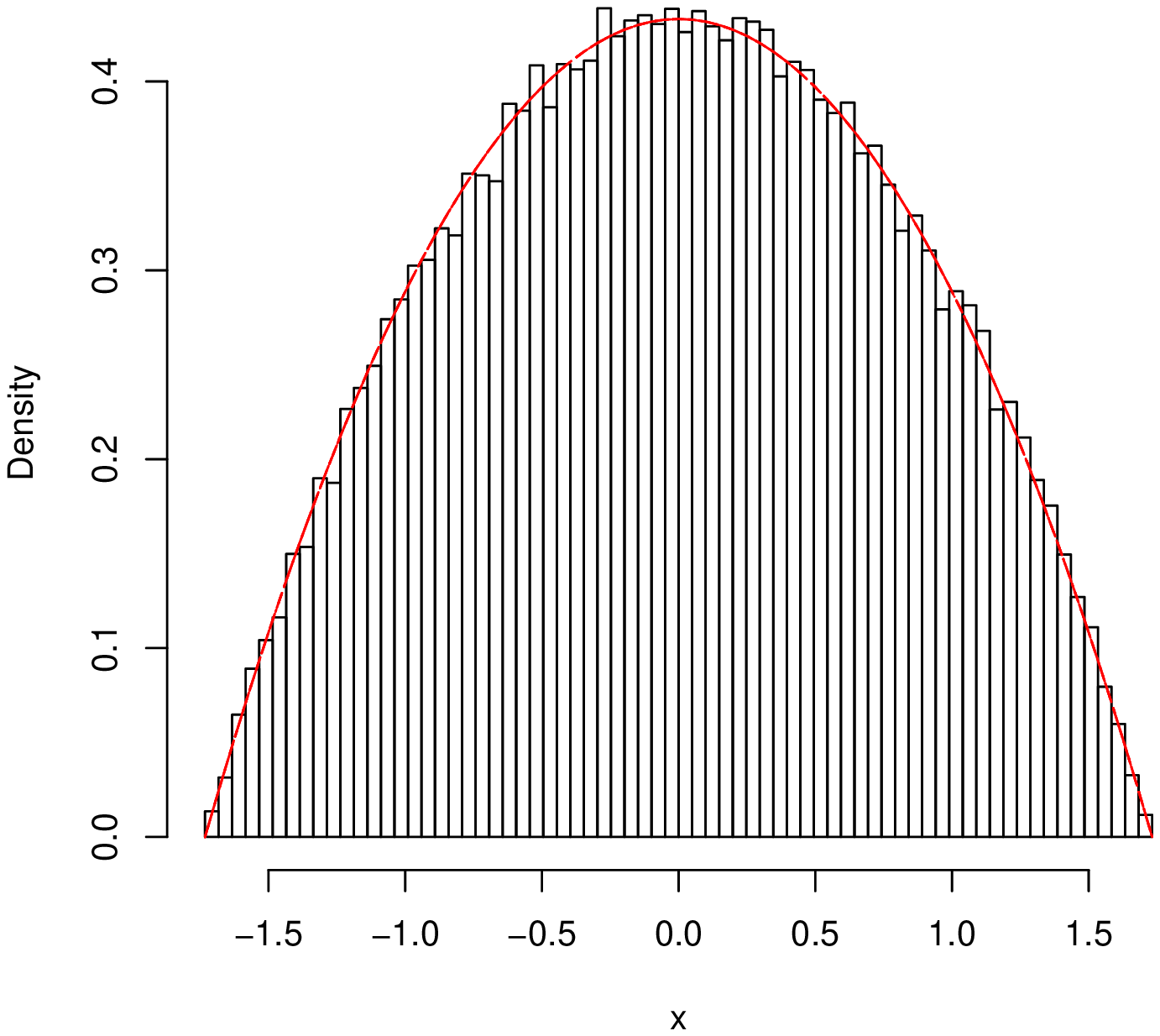}
		\caption{Histogram and $q$PDF for $q=0$}
		\label{fig:sub2}
	\end{subfigure}
	\caption{The standard q-Gaussian distribution shapes for representatives values of $q$, and a histogram with rqgauss(n, q = 0)}		
	\label{figure:2}
\end{figure}

In the next example, we compare $q$-Gaussian against two particular cases.
The $q$-Gaussian is related with Student's-t distribution by $q=(3+\text {df})/(1+\text {df})$ and Cauchy ($q=2$) \cite{TS97}
 and  \cite{AN05}. At theses codes, first we can seen how  Student's-t distribution is a particular case of a more general distribution, using the standard $q$PDF as model. We generate a random 
sample by means R \textbf{stats} package using the command {\fontfamily{cmss}\selectfont{rt(n,df)}} with length {\fontfamily{cmss}\selectfont{n}} for {\fontfamily{cmss}\selectfont{df}}
degree of freedom. At second, the cumulative Cauchy distribution is presented versus {\fontfamily{cmss}\selectfont{rqgauss}} random number generator.

{\fontfamily{cmss}\selectfont
\noindent
$\#\#\#$ qGaussian versus Student-t\\
set.seed(1234)\\
sam <- 1000; df <- 7\\
r <- rt(sam, df)\\
qv <- (df+3)/(df+1)\\
plot(sort(r), (1:sam/sam), main = "qCDF vs rt", col = "blue",\\
	ylab =  "Probability", xlim = range(-4.5, 4.5), xlab = 'x')\\
x <-  seq(min(r), max(r), length = 313)\\
lines(x, pqgauss(x, qv), lwd = 2)\\
legend(1.5, 0.7, legend = c(expression(paste(q == 1.25)), expression(paste(df == 7))),\\
	col = c("black", "blue"), lty = c(1, 0), lwd = 1, pch = c(-1, 1))\\
\noindent		
$\#\#\#$ qGaussian versus Cauchy\\
set.seed(1234)\\
sam <- 1000\\
r2 <- rcauchy(sam, 100)\\
x2 <- 1:sam/sam\\
plot(x2, sort(r2), main = "qQF vs rcauchy", col = "red",\\
	xlab = "Probability", ylim = range(70, 160), ylab = 'x')\\
lines(x2, cqgauss(x2, mu = 100, q = 2), lwd = 2)\\
legend(.3, 145, col = c("black", "red"), lty = c(1, 0), lwd = 1, pch = c(-1, 1),\\
	legend = c(expression(paste('q == 2')), expression(paste('rcauchy(n, 100)')))) \\ 
}

\begin{figure}[htbp]
  \centering
  \begin{subfigure}{.5\textwidth}
  	\centering
  	\includegraphics[width=1\linewidth]{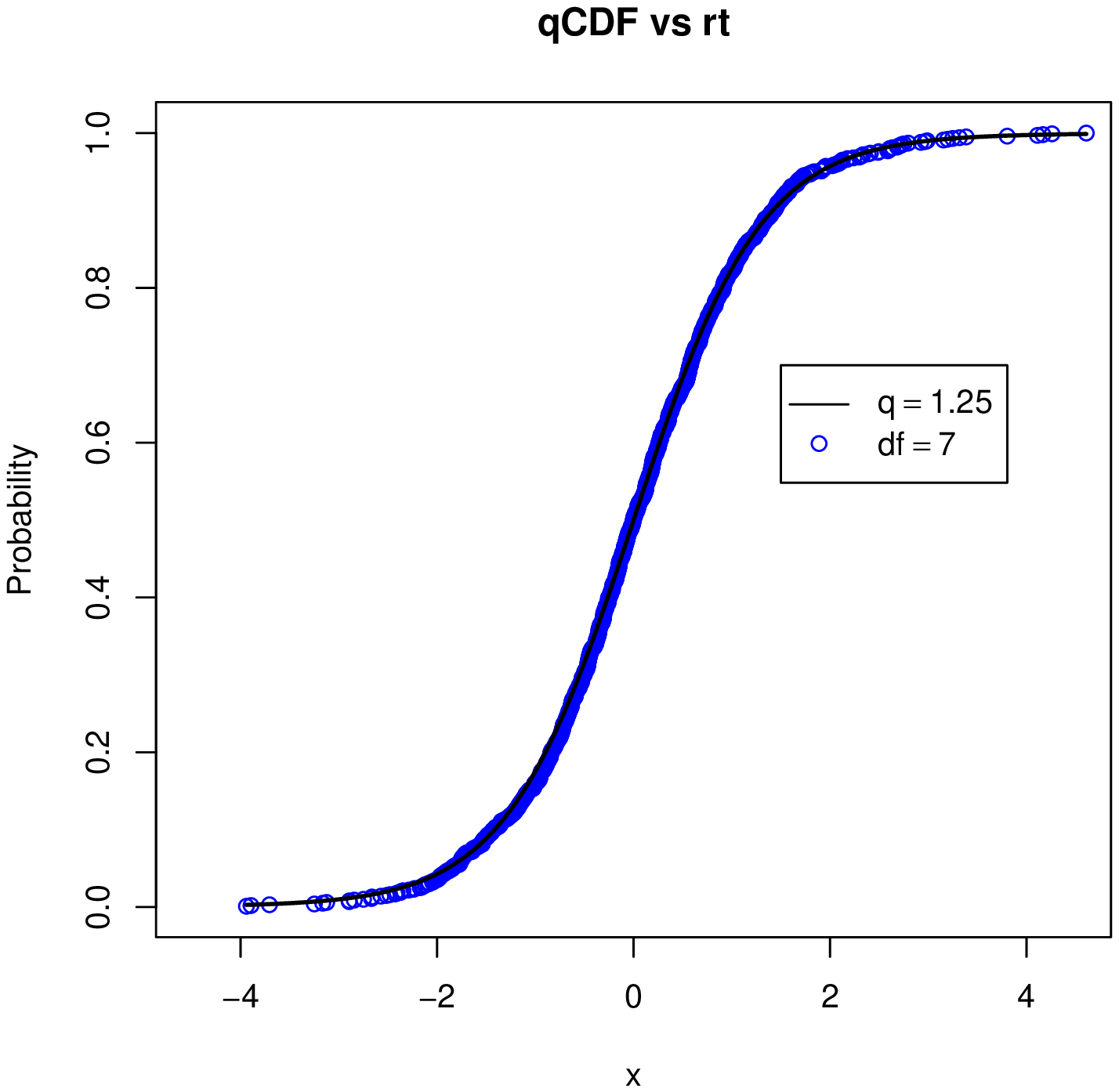}
  	\caption{\centering{The blue dots was created from
  			Student's-t random generator,{\fontfamily{cmss}\selectfont{rt(..)}}, and fitted with a $q$CDF}}
  	\label{fig:sub3}
  \end{subfigure}%
  \begin{subfigure}{.5\textwidth}
  	\centering
  	\includegraphics[width=1.\linewidth]{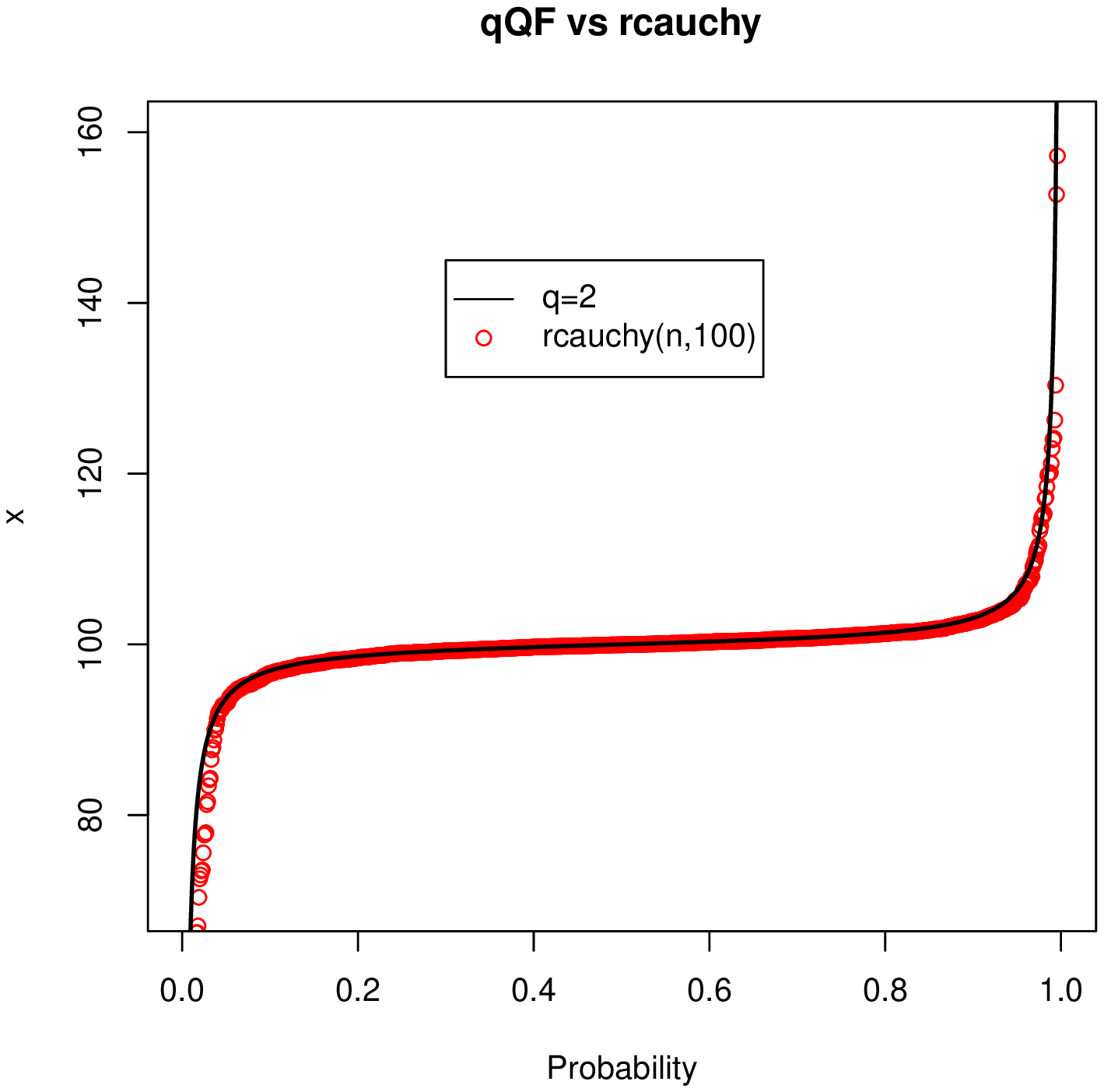}
  	\caption{\centering{The solid line, $q$QF, matches with the red dots generated using {\fontfamily{cmss}\selectfont{rcauchy}} command.}}
  	\label{fig:sub4}
  \end{subfigure}
  \caption{Comparison of $q$-Gaussian against two specials cases}
 \label{figure:comp}
\end{figure}

In the third example, we will figure out how estimate the $q$ value for a random sample using the {\fontfamily{cmss}\selectfont{qbymc(X)}} command. For that, we generate a synthetic data with parameters $q=1.39$ and length $n=2004$. After, it is shown that the value of $q$ and its standard error that went obtained remain unchanged, regardless of the values chosen for $q$-mean and $q$-variance.

{\fontfamily{cmss}\selectfont
\noindent
$\#\#\#$ Identifying a random sample\\
set.seed(1000)\\
qbymc(rqgauss(2004, 1.39)) \\
}
{\fontfamily{cmss}\selectfont
\noindent
Estimate Std. Error  \\ 
1.411094   0.109256  \\
}

{\fontfamily{cmss}\selectfont
\noindent
$\#\#\#$ Identifying a random sample regardless q and mu \\
set.seed(1000)\\
qbymc(rqgauss(2004, 1.39, 3.141592, 2.718281))\\
}
{\fontfamily{cmss}\selectfont
\noindent
Estimate Std. Error \\
1.411094   0.109256 \\
}

\section{Summary}

In this work, we create a statistical package for $q$-Gaussian distribution following the pattern of R \textbf{stats} packages. Also, was included an algorithm that is used to identify $q$PDF at an empirical data set. Moreover, we hope to include in future releases, other mathematical topics related to $q$ algebra and
others nonextensive distributions.


\bibliography{ref}

\begin{thebibliography}{10}
\expandafter\ifx\csname url\endcsname\relax
  \def\url#1{\texttt{#1}}\fi
\expandafter\ifx\csname urlprefix\endcsname\relax\def\urlprefix{URL }\fi
\expandafter\ifx\csname href\endcsname\relax
  \def\href#1#2{#2} \def\path#1{#1}\fi

\bibitem{T88}
C.~Tsallis, Possible generalization of boltzmann-gibbs statistics, J. Stat.
  Phys. 52~(1/2) (1988) 479.

\bibitem{T09}
C.~Tsallis, Introduction to Nonextensive Statistical Mechanics, Springer, 2009.

\bibitem{H14}
A.~Sato, Applied Data-Centric Social Sciences: Concepts, Data, Computation, and
  Theory, Springer-Verlag, 2014.

\bibitem{GT04}
M.~Gell-Mann, C.~Tsallis (Eds.), Nonextensive Entropy—Interdisciplinary
  Applications, Oxford University Press, 2004.

\bibitem{SH15}
E.~L. de~Santa~Helena, C.~M. Nascimento, G.~J. Gerhardt, Alternative way to
  characterize a q-gaussian distribution by a robust heavy tail measurement,
  Physica A 435~(1) (2015) 44--50.

\bibitem{AS83}
M.~Abramowitz, I.~A. Stegun, Handbook of Mathematical Functions with Formulas,
  Graphs, and Mathematical Tables. Applied Mathematics Series, Dover
  Publications, 1983.

\bibitem{TNT07}
W.~Thistleton, J.~A. Marsh, K.~Nelson, C.~Tsallis, IEEE Transactions on
  Information Theory 53(12) (2007) 4805.

\bibitem{BHS06}
G.~Brys, M.~Hubert, A.~Struyf, Comput. Statist. Data Anal. 50 (2006) 733.

\bibitem{TS97}
A.~de~Souza, C.~Tsallis, Physica A 236 (1997) 52--57.

\bibitem{AN05}
C.~Anteneodo, Physica A 358 (2005) 289--298.

\end{thebibliography}

\end{document}